\documentclass[prl,twocolumn,nofootinbib]{revtex4}
\usepackage{amsmath,amssymb}
\usepackage{graphicx}

\begin{document}

\vspace{6.0cm}

\title{Renormalizable Adjoint SU(5)}

\vspace{0.3cm}

\author{Pavel Fileviez P\'erez}

\vspace{0.5cm}

\email{fileviez@cftp.ist.utl.pt}

\affiliation{
Centro de F{\'\i}sica Te\'orica de Part{\'\i}culas 
\\
Departamento de F{\'\i}sica.\ Instituto Superior T\'ecnico 
\\
Avenida Rovisco Pais 1, 1049-001 Lisboa, Portugal 
\\
and
\\
Max Planck Institut f\"ur Physik
\\
(Werner-Heisenberg-Institut)
\\
F\"ohringer Ring 6. D-80805 M\"unchen, Germany}
\begin{abstract}
We investigate the possibility to find the simplest renormalizable 
grand unified theory based on the SU(5) gauge symmetry. We find that 
it is possible to generate all fermion masses with only two 
Higgs bosons, $5_H$ and $45_H$. In this context the neutrino masses 
are generated through the type III and type I seesaw mechanisms.
The predictions coming from the unification of gauge couplings and 
the stability of the proton are discussed in detail. In this 
theory the leptogenesis mechanism can be realized through the out of 
equilibrium decays of the fermions in the adjoint representation.
\end{abstract}
\maketitle
{\section{I. Introduction}} 

The possibility to unify all fundamental interactions in nature is 
one of the main motivations for physics beyond the Standard Model (SM). 
The so-called grand unified theories (GUTs) are considered as one of 
the most natural extensions of the Standard Model where this dream 
is partially realized. Two generic predictions of those theories are 
the unification of gauge interactions at the high scale, 
$M_{GUT} \approx 10^{14- 16}$ GeV, and the decay of the lightest 
baryon~\cite{Pati-Salam}, the proton, which unfortunately still 
has not been observed in the experiments.

The first grand unified theory was proposed by Georgi and 
Glashow in reference~\cite{GG}. As is well known this model, 
based on $SU(5)$ gauge symmetry, has been considered as the 
simplest grand unified theory. It offers partial matter unification 
of one Standard Model family in the anti-fundamental ${\overline{5}}$ 
and antisymmetric $10$ representations. The Higgs sector is 
composed of ${24}_H$ and ${5}_H$. The GUT symmetry is broken down 
to the Standard Model by the vacuum expectation value (VEV) of the 
Higgs singlet field in ${24}_H$, while the SM Higgs resides in ${5}_H$. 
The beauty of the model is undeniable, but the model itself is not 
realistic. This model is ruled out for three reasons: the unification 
of the gauge couplings is in disagreement with the values 
of $\alpha_{em}$, $\sin^2 \theta_W$ and $\alpha_s$ at the electroweak 
scale, the neutrinos are massless and the unification of 
the Yukawa couplings of charged leptons and down quarks at the 
high scale in the renormalizable model 
is in disagreement with the experiments.

Recently, several efforts has been made in order to define the
simplest realistic extension of the Georgi-Glashow model. 
The simplest realistic grand unified theory with the Standard 
Model matter content was pointed out in reference~\cite{Ilja-Pavel} 
where the ${15}_H$ has been used to generate neutrino masses 
and achieve unification. For different phenomenological and 
cosmological aspects of this proposal see references~\cite{IPR}, 
\cite{IPG}, \cite{I}, and \cite{P}. This theory predicts for the 
first time the existence of light scalar leptoquarks and 
that the upper bound on the proton lifetime is 
$\tau_p \lesssim 2 \times 10^{36}$ years. Therefore, this realistic 
grand unified theory could be tested at future collider experiments, 
particularly at LHC, through the production of scalar leptoquarks 
and at next generation of proton decay experiments. 
Now, if we extend the Georgi-Glashow model adding extra matter, 
there is a realistic grand unified model where the extra matter 
is in the ${24}$ representation. This possibility has been proposed 
recently by Bajc and Senjanovi\'c in reference~\cite{Borut-Goran}. 
In this scenario using higher-dimensional operators the neutrino masses 
are generated through the type I~\cite{TypeI} and type III~\cite{TypeIII} 
seesaw mechanisms. In this case the theory predicts a light fermionic 
$SU(2)_L$ triplet~\cite{Borut-Goran} which is responsible for type III seesaw. 
See references~\cite{Borut-Goran}, and \cite{Ilja-Pavel-24F} for more details. 

The type III seesaw mechanism has been proposed for the first time in 
reference~\cite{TypeIII}. In this case adding at least two fermionic 
$SU(2)_L$ triplets with zero $U(1)_Y$ hypercharge the effective dimension 
five operator relevant for neutrino masses are generated once the neutral 
components of the fermionic triplets are integrated out~\cite{TypeIII}.
In the context of grand unification Ma studied for the first time 
the implementation of this mechanism in SUSY $SU(5)$~\cite{Ma}.
In this case a fermionic chiral matter superfield in the $24$ 
representation has to be introduced~\cite{Ma} and the neutrino masses 
are generated through type I and type III seesaw mechanisms since in 
the $24$ representation one has the fermionic triplet responsible 
for type III seesaw and a singlet responsible for type I seesaw. 
Therefore, if we want to realize the type III seesaw mechanism 
one must introduce extra matter in the adjoint representation.
The implementation of this mechanism in non-SUSY $SU(5)$ has been 
understood in reference~\cite{Borut-Goran}. In this case they 
have introduced extra matter in the $24$ representation and 
use higher-dimensional operators in order to generate at least 
two massive neutrinos and a consistent relation between the masses 
of charged leptons and down quarks~\cite{Borut-Goran}.

The models mentioned above include the whole set of higher 
dimensional operators in order to have a consistent relation 
between the Yukawa couplings at the unification scale. 
In this work we want to stick to the renormalizability principle 
and focus our attention on renormalizable extensions of the 
Georgi-Gashow model. Following the results presented in 
references~\cite{Ma} and~\cite{Borut-Goran} we investigate 
the possibility to write down the simplest renormalizable 
grand unified theory based on the $SU(5)$ gauge symmetry. 
We find that it is possible to generate all fermion masses at the 
renormalizable level , including the neutrino masses, 
with the minimal number of Higgs bosons: $5_H$ and $45_H$. 
The implementation of the leptogenesis mechanism
~\cite{leptogenesis} is possible. In this model 
the leptogenesis mechanism can be realized through 
the out of equilibrium decays of the fermions in 
the adjoint representation. Notice that in the model proposed 
in reference~\cite{Borut-Goran} only resonant leptogenesis 
could be possible since the fermionic triplet is very light. 
As we will show in the next section there is no problem to 
satisfy the experimental lower bounds on the proton decay 
lifetime. We propose a new renormalizable grand unified 
theory based on the $SU(5)$ gauge symmetry with extra 
matter in the adjoint representation. We refer to this theory 
as ``Renormalizable Adjoint $SU(5)$''. The model 
proposed in this letter can be considered as the renormalizable 
version of the model given in reference~\cite{Borut-Goran} 
and is one of the most appealing candidates for the unification 
of the Standard Model interactions at the renormalizable level. 
In the next sections we discuss some of the most relevant 
phenomenological aspects of this proposal. 
\\
{ \section{ II. Renormalizable Adjoint $SU(5)$}} 

In order to write down a realistic grand unified theory we have to be sure 
that all constraints coming from the unification of gauge couplings, 
fermion masses and proton decay can be satisfied. In this letter we stick 
to the simplest unified gauge group, $SU(5)$, and to the renormalizability 
principle. Now, if we want to have a consistent relation between the masses of 
charged leptons and down quarks at the renormalizable level we have to 
introduce the $45_H$ representation~\cite{Georgi45}. Therefore, our Higgs sector 
must be composed of ${{24}}_H=(\Sigma_8,\Sigma_3,\Sigma_{(3,2)},\Sigma_{(\bar{3}, 2)},\Sigma_{24})$
$=({8},{1},0)\bigoplus({1},{3},0)\bigoplus ({3},{2},-5/6)\bigoplus(\overline{{3}},{2},5/6)
\bigoplus ({1},{1},0)$,  
${45_H}=(\Phi_1, \Phi_2, \Phi_3, \Phi_4, \Phi_5, \Phi_6, H_2)=({8},{2},1/2)\bigoplus$
$(\overline{{6}},{1},-1/3)\bigoplus({3},{3},-1/3)\bigoplus(\overline{{3}},{2},-7/6)
\bigoplus({3},{1},-1/3)\bigoplus$
$(\overline{{3}},{1},4/3)\bigoplus({1}, {2},1/2)$, and ${{5}}_H=(H_1,T)=({1},{2},1/2)$
$\bigoplus({3},{1},-1/3)$ where the field ${45}$ satisfies the following conditions: 
$({45})^{\alpha \beta}_{\delta} = - ({45})^{\beta \alpha}_{\delta}$, 
$\sum_{\alpha=1}^5 ({45})^{\alpha \beta}_{\alpha} = 0$, and 
$v_{45} = \langle 45 \rangle^{1 5}_{1}= \langle 45 \rangle^{2 5}_{2}= \langle 45 \rangle^{3 5}_{3}$. 
In this model the Yukawa potential for charged fermions reads as:
\begin{eqnarray}
V_{Y} &=& {10} \ \overline{{5}} \ \left( Y_1 \ {5}^*_H \ + \  Y_2 \ {45}^*_H \right) \ + \nonumber\\
&+& {10} \ {10} \ \left( Y_3 \ {5_H} \ + \ Y_4 \ 45_H \right) \ + \ h.c. 
\end{eqnarray}
and the masses for charged leptons and down quarks are given by:
\begin{eqnarray}
M_D &=& Y_1 \ v_5^* \ + \ 2 Y_2 \ v_{45}^* ,\\
M_E &=& Y_1^T \ v_5^* \ - 6 \ Y_2^T \ v_{45}^* , \label{GJ}
\end{eqnarray}
where $\langle{5}_H \rangle=v_5$. $Y_1$ and $Y_2$ 
are arbitrary $3 \times 3$ matrices. Notice that there are 
clearly enough parameters in the Yukawa sector to fit 
all charged fermions masses. See reference~\cite{potential} 
for the study of the scalar potential and~\cite{IPG} for the 
relation between the fermion masses at the high scale which is 
in agreement with the experiment. 

There are three different possibilities to generate the neutrino 
masses~\cite{Ma} at tree level in this context. The model can be 
extended in three different ways: i) we can add at least two 
fermionic $SU(5)$ singlets and generate neutrino masses 
through the type I seesaw mechanism~\cite{TypeI}, ii) we can add 
a $15$ of Higgs and use the type II seesaw~\cite{TypeII} mechanism, 
or iii) we can generate neutrino masses through the type III~\cite{TypeIII} 
and type I seesaw mechanisms adding at least two extra matter fields 
in the $24$ representation~\cite{Ma}. In reference~\cite{Borut-Goran} it has 
been realized the possibility to generated the neutrino masses through 
type III and type I seesaw adding just one extra matter field in $24$ and using 
higher-dimensional operators. Notice that the third possibility mentioned 
above is very appealing since we do not have to introduce 
$SU(5)$ singlets or an extra Higgs. If we add an extra 
Higgs, $15_H$, for type II seesaw mechanism the Higgs sector 
is even more complicated. In this letter we focus on the 
possibility to generate the neutrino masses at the renormalizable 
level through type III and type I seesaw mechanisms.  

The predictions coming from the unification of the gauge couplings 
in a renormalizable $SU(5)$ model where one uses type I or type II 
seesaw mechanism for neutrino masses were investigated 
in reference~\cite{Ilja-Pavel-45}. However, a renormalizable 
grand unified theory based on $SU(5)$ where the neutrino masses 
are generated through the type III seesaw mechanism has not been 
proposed and this is our main task. The SM decomposition of the needed \
extra multiplet for type III seesaw is given by: 
$
{24}= (\rho_8,\rho_3, \rho_{(3,2)}, \rho_{(\bar{3}, 2)},
\rho_{0})=({8},{1},0)\bigoplus({1},{3},0)\bigoplus({3},{2},-5/6)
\bigoplus(\overline{{3}},{2},5/6)\bigoplus({1},{1},0) 
$. 
In our notation $\rho_3$ and $\rho_0$ are the $SU(2)_L$ 
triplet responsible for type III seesaw and the singlet 
responsible for type I seesaw, respectively. Since we have 
introduced an extra Higgs $45_H$ and an extra matter multiplet $24$, 
the Higgs sector of our model is composed of $5_H$, $24_H$ and $45_H$, 
and the matter is unified in the $\bar{5}$, $10$ and $24$ representations. 

The new relevant interactions for neutrino masses in this context 
are given by:

\begin{equation}
\label{neutpot}
V_\nu = c_i \ \overline{5}_i \ 24 \ 5_H \ + \ p_i \ \overline{5}_i \ 24 \ 45_H  \ + \ h.c. 
\end{equation}      

Notice from Eq.~(1) and Eq.~(4) the possibility to generate all fermion masses, 
including the neutrino masses, with only two Higgses : $5_H$ and $45_H$. The first term 
in the above equation has been used in reference~\cite{Ma} in the context of 
SUSY $SU(5)$ and in reference~\cite{Borut-Goran} in the context of non-SUSY 
$SU(5)$. Notice that the main difference at this level of our model with the model 
presented in reference~\cite{Borut-Goran} is that we do not need to use 
higher-dimensional operators and with only two Higgses we can generate 
all fermion masses. Notice that in $SU(5)$ models usually that is not possible.

Using Eq.~(\ref{neutpot}) the neutrino mass matrix reads as:

\begin{eqnarray}
M^\nu_{ij} & = & \frac{a_i a_j}{M_{\rho_3}} \ + \ \frac{b_i b_j}{M_{\rho_0}}, 
\end{eqnarray} 
where
\begin{equation}
a_i = c_i  v_5 \ - \  3 p_i v_{45}, 
\end{equation}
and
\begin{equation}
b_i  =  \frac{\sqrt{15}}{2} \left( \frac{c_i  v_5}{5} \ + \ p_i v_{45}  \right). 
\end{equation}   
The theory predicts one massless neutrino at tree level. Therefore, we could 
have a normal neutrino mass hierarchy: $m_1=0$, $m_2=\sqrt{\Delta m_{sun}^2}$ 
and $m_3=\sqrt{\Delta m_{sun}^2 + \Delta m_{atm}^2}$ or the inverted neutrino 
mass hierarchy: $m_3=0$, $m_2=\sqrt{\Delta m_{atm}^2}$ and 
$m_1=\sqrt{\Delta m_{atm}^2 - \Delta m_{sun}^2}$. $\Delta m_{sun}^2 \approx 8 \times 10^{-5}$ eV$^{2}$ 
and $\Delta m_{atm}^2 \approx 2.5 \times 10^{-3}$ eV$^{2}$ are the mass-squared 
differences of solar and atmospheric neutrino oscillations~\cite{PDG}, respectively. 

The masses of the fields responsible for the seesaw mechanisms are computed 
using the new interactions between $24$ and $24_H$ in this model:
\begin{equation}
V_{24} = m \ Tr (24^2) \ + \ \lambda \ Tr (24^2 24_H)  
\end{equation}
Once $24_H$ gets the expectation value, $\langle 24_H \rangle = v \ diag (2,2,2, -3, -3) / \sqrt{30}$, 
the masses of the fields living in $24$ are given by: 
\begin{eqnarray}
M_{\rho_0} &=& m -\frac{\tilde{\lambda} M_{GUT}}{\sqrt{\alpha_{GUT}}},\\
M_{\rho_3} &=& m -\frac{3 \tilde{\lambda} M_{GUT}}{\sqrt{\alpha_{GUT}}},\\
M_{\rho_8} &=& m + \frac{2 \tilde{\lambda} M_{GUT}}{\sqrt{\alpha_{GUT}}},\\
M_{\rho_{(3,2)}} &=& M_{\rho_{(\bar{3},2)}} = m -\frac{\tilde{\lambda} M_{GUT}}{2 \sqrt{\alpha_{GUT}}},
\end{eqnarray}
where we have used the relations $M_V= v \sqrt{5 \pi \alpha_{GUT}/3}$, 
$\tilde{\lambda}= \lambda / {\sqrt{50 \pi}}$ and chose $M_V$ as the 
unification scale. Notice that when the fermionic triplet $\rho_3$, 
responsible for type III seesaw mechanism, is very light the rest 
of the fields living in $24$ have to be heavy if we do not assume 
a very small value for the $\lambda$ parameter.  

Before study the unification constraints and discuss the 
different contributions to proton decay let us summarize our results.
We have found that it is possible to write down a renormalizable 
non-supersymmetric grand unified theory based on the $SU(5)$ gauge symmetry 
where the neutrino masses are generated through type I and type III 
seesaw mechanisms using just two Higgses $5_H$ and $45_H$. In this context, 
as in the model proposed in reference~\cite{Borut-Goran}, the implementation 
of leptogenesis is possible. However, in their case one could 
have only resonant leptogenesis. Those issues will be discussed in detail 
in a future publication.
\\
{\section{III. Unification Constraints and Nucleon Decay }} 

In order to understand the constraints coming from the unification 
of gauge couplings we can use the B-test relations: 
$ B_{23}/ B_{12}= 0.716 \pm 0.005$ and $\ln  M_{GUT}/ {M_Z}= ( 184.9 \pm 0.2 ) / B_{12}$, 
where the coefficients $B_{ij}=B_i-B_j$ and $B_i = b_i + \sum_{I} b_{iI} \ r_{I}$ are 
the so-called effective coefficients. Here $b_{iI}$ are the appropriate one-loop 
coefficients of the particle $I$ and $r_I=(\ln M_{GUT}/M_{I})/(\ln M_{GUT}/M_{Z})$ 
($0 \leq r_I \leq 1$) is its ``running weight''~\cite{Hall}. To obtain the above 
expressions we have used the following experimental values at $M_Z$ in 
the $\overline{MS}$ scheme~\cite{PDG}: $\sin^2 \theta_W (M_Z)=0.23120 \pm 0.00015$, 
$\alpha_{em}^{-1}(M_Z)=127.906 \pm 0.019$ and $\alpha_{s}(M_Z)=0.1176 \pm 0.002$. 
In the rest of the paper we will use the central values for input parameters 
in order to understand the possible predictions coming from the unification 
of gauge interactions.  

As is well known the $B$-test fails badly in the Standard Model case since 
$B_{23}^{SM}/B_{12}^{SM}=0.53$, and hence the need for extra light particles 
with suitable $B_{ij}$ coefficients to bring the value of the $B_{23}/B_{12}$ 
ratio in agreement with its experimental value. In order to understand this 
issue we compute and list the $B_{ij}$ coefficients of the different fields 
in our model in Tables~\ref{table1}, \ref{table2}, and \ref{table3}. 
Notice that we have chosen the mass of the superheavy gauge bosons as the 
unification scale. From the tables we see clearly that $\Sigma_3$, $\Phi_3$ 
and $\rho_3$ fields improve unification with respect to the Standard Model 
case since those fields have a negative and positive contribution 
to the coefficients $B_{12}$ and $B_{23}$, respectively. 

\begin{table}[h]
\caption{\label{table1} Contributions of $5_H$, and $24_H$ multiplets 
to the $B_{ij}$ coefficients, including the contribution of the Higgs 
doublet in $45_H$. The masses of the Higgs doublets are taken to be at $M_Z$.}
\begin{ruledtabular}
\begin{tabular}{lcccccc}
 & 2HSM & $T$ & $\Sigma_8$ & $\Sigma_3$ \\
\hline $B_{23}$ & $4$ & $-\frac{1}{6} r_{T}$ & $-\frac{1}{2} r_{\Sigma_8}$ & $\frac{1}{3} r_{\Sigma_3}$ \\
$B_{12}$ & $\frac{36}{5}$ & $\frac{1}{15} r_{T}$ & 0 & $-\frac{1}{3} r_{\Sigma_3}$
\end{tabular}
\end{ruledtabular}
\end{table}
\begin{table}[h]
\caption{\label{table2} Contributions of the fields in $45_H$ 
to the $B_{ij}$ coefficients, excluding the contribution of the 
Higgs doublet $H_2$.}
\begin{ruledtabular}
\begin{tabular}{lccccccc}
 & $\Phi_1$ & $\Phi_2$ & $\Phi_3$ & $\Phi_4$ & $\Phi_5$ & $\Phi_6$\\
\hline $B_{23}$ & $-\frac{2}{3}r_{\Phi_1}$ & $- \frac{5}{6} r_{\Phi_2}$ &
$\frac{3}{2} r_{\Phi_3}$ & $\frac{1}{6} r_{\Phi_4}$
& $-\frac{1}{6} r_{\Phi_5}$ &  $-\frac{1}{6} r_{\Phi_6}$\\
$B_{12}$ & $-\frac{8}{15} r_{\Phi_1}$ & $\frac{2}{15} r_{\Phi_2}$ & $-\frac{9}{5} r_{\Phi_3}$
& $\frac{17}{15} r_{\Phi_4}$ & $\frac{1}{15} r_{\Phi_5}$ & $\frac{16}{15} r_{\Phi_6}$
\end{tabular}
\end{ruledtabular}
\end{table}
\begin{table}[h]
\caption{\label{table3} Extra contributions of the extra matter 
in the multiplet $24$ to $B_{ij}$ coefficients.}
\begin{ruledtabular}
\begin{tabular}{lccccccccc}
     & $\rho_8$ & $\rho_3$ & $\rho_{(3,2)}$ & $\rho_{(\bar{3},2)}$ \\
\hline $B_{23}$& $- 2 r_{\rho_8}$ & $\frac{4}{3}r_{\rho_3}$ &
$\frac{1}{3} r_{\rho_{(3,2)}}$ & $\frac{1}{3} r_{\rho_{(\bar{3},2)}}$\\
$B_{12}$&  0  & $- \frac{4}{3}r_{\rho_3}$ &
$\frac{2}{3} r_{\rho_{(3,2)}}$ & $\frac{2}{3} r_{\rho_{(\bar{3},2)}}$
\end{tabular}
\end{ruledtabular}
\end{table}
Before we study the different scenarios in agreement with the 
unification of gauge interactions let us discuss the different 
contributions to proton decay. For a review on proton decay 
see~\cite{review}. In this model there are five multiplets 
that mediate proton decay. These are the superheavy gauge bosons
$V=({3},{2},-5/6)\bigoplus(\overline{{3}},{2},5/6)$, the $SU(3)$ 
triplet $T$, $\Phi_3$, $\Phi_5$ and $\Phi_6$. The least
model dependent and usually the dominant proton decay
contribution in non-supersymmetric scenarios comes from gauge 
boson mediation. Its strength is set by $M_V$ and 
$\alpha_{GUT}$. Notice that we have identified $M_V$ 
with the GUT scale, i.e., we set $M_V \equiv M_{GUT}$. 
We are clearly interested in the regime where $M_V$ 
is above the experimentally established bounds set by proton decay, 
$M_V \gtrsim (2 \times 10^{15}) \ 5 \times 10^{13}$ GeV if 
we do (not) neglect the fermion mixings~\cite{upper}.    

In this theory the value of $M_{GUT}$ depends primarily on the masses 
of $\Sigma_3$, $\rho_3$, $\Phi_1$ and $\Phi_3$ through their 
negative contributions to the $B_{12}$ coefficient. 
The $\Phi_3$ field cannot be very light due 
to proton decay constraints. The $\Phi_3$ 
contributions to proton decay are coming from interactions 
$Y_4 Q^T \imath \sigma_2 \Phi_3 Q$ and 
$Y_2 Q^T \imath \sigma_2 \Phi_3^* L$. The field 
$\Phi_3$ should be heavier than $10^{11}$\,GeV in order 
to not conflict experimental data. Of course, this rather 
naive estimate holds if one assumes most natural values 
for Yukawa couplings. If for some reasons one of the two 
couplings is absent or suppressed the bound on $\Phi_3$ 
would cease to exist. For example, if we choose $Y_4$ to be 
an anti-symmetric matrix, the coupling $Y_4 Q^T \imath \sigma_2 \Phi_3 Q$ 
vanishes. Therefore, $\Phi_3$ could be very light. 
In general the field $\Sigma_3$ could be between the electroweak 
and the GUT scales, while $\rho_3$ has to be always below the 
seesaw scale, $M_{\rho_3} \lesssim 10^{14}$ GeV.  
Let us study several scenarios where the 
unification constraints are quite different:

\begin{itemize}
\item 
The first scenario corresponds to the case when 
$\Sigma_3$ is at GUT scale, while $\Phi_3$ 
and/or $\rho_3$ could be below the unification 
scale. The rest of the fields are at the unification scale. 
Using the $B_{ij}$ coefficients listed in Tables.~I--III 
we find that if $\Phi_3$ is at GUT scale it is possible 
to achieve unification at $1.83 \times 10^{14}$ GeV 
if $M_{\rho_3}=1.13 \times 10^{8}$ GeV.  
However, if $\rho_3$ is at the seesaw scale, $10^{14}$ GeV, we 
achieve unification at $2.46 \times 10^{14}$ GeV if 
$M_{\Phi_3}=3.68 \times 10^{9}$ GeV. In both cases 
the unification scale is rather low and it is possible to 
achieve unification with only one of these fields, $\Phi_3$ or $\rho_3$, 
since in this model one can have two light Higgs doublets at $M_Z$.
Notice that in the first case we have to suppress the gauge 
contributions to nucleon decay, while in the second 
case both the $\Phi_3$ and gauge contributions have to be suppressed 
in order to satisfy the experimental bounds on proton decay lifetimes, 
typically $\tau_p^{exp} \gtrsim 10^{33}$ years.  
\item 
In the second scenario $\Sigma_3$ is at the electroweak scale. 
In this case if $\Phi_3$ is at the GUT scale and $M_{\rho_3}=1.35 \times 10^{11}$ 
GeV the gauge couplings unify at $1.83 \times 10^{14}$ GeV. Now, in the case 
when $\rho_3$ is at the seesaw scale the unification is at $2.11 \times 10^{14}$ 
GeV if $M_{\Phi_3}=1.01 \times 10^{12}$ GeV. Notice that as in the previous 
scenario the unification scale is very low, while the mass of $\Phi_3$ is 
always above the lower bound coming from nucleon decay. Therefore, we only 
have to suppress the gauge contributions to proton decay through the fermionic 
mixings~\cite{upper}.      
\item 
In the previous scenarios we have assumed that the fields $\Phi_3$, 
$\rho_3$ and $\Sigma_3$ could be below the GUT scale, while the rest of 
the fields are at the unification scale. Now let us analyze the case when 
those fields can contribute to the running of gauge couplings. 
In particular the contributions of $\Phi_1$ and $\Sigma_8$ are quite relevant 
in order to understand what is the maximal unification scale in our model. 
Notice that $\Phi_1$ has negative contributions to the $B_{23}$ and 
$B_{12}$ coefficients, while $\Sigma_8$ has only negative contribution to $B_{23}$. 
When those fields, $\Phi_1$ and $\Sigma_8$, are very light the unification 
scale will be higher than in the previous scenarios since in this case 
the rest of the fields have to be lighter in order to satisfy the B-test relations.    
It is easy to understand that the maximal unification scale in this scenario 
corresponds to the case when $M_{\Phi_1}=M_{\Sigma_8}=M_Z$, $M_{\Sigma_3}=M_{GUT}$, 
$M_{\Phi_3} = 1.2 \times 10^{9}$ GeV and $M_{\rho_3}=10^{14}$ 
GeV. In this case the unification scale is $M_{GUT}=1.2 \times 10^{17}$ GeV.  
Therefore, in this case one can conclude that there is no hope to 
test this scenario at future proton decay experiments since 
$\tau_p \gtrsim 10^{41}$ years. 
\end{itemize}
It is important to know which is the minimal value for the mass 
of the fermionic triplet, responsible for type III seesaw, consistent 
with unification. The minimal value of $M_{\rho_3}$ corresponds to 
the case when $\Sigma_3$ and $\Phi_3$ are at the GUT scale, 
while $\Phi_1$ and $\Sigma_8$ are close to the electroweak scale. 
In this case $M_{\rho_3} \approx 1.5$ TeV 
and the unification scale is $3 \times 10^{16}$ GeV. 
Therefore, we can conclude that in this case the seesaw mechanism 
could be tested at future collider experiments. 
The minimal value of $M_{\Phi_3}$ in our model is $5 \times 10^8$ GeV 
when $M_{\Phi_1}=M_{\Sigma_8}\approx M_Z$,
$M_{\Sigma_3}\approx M_{GUT} \approx 5 \times 10^{16}$ GeV and 
$M_{\rho_3}=10^{14}$ GeV. Notice that in all the scenarios 
studied in this section we can satisfy the constraints coming from 
proton decay and neutrino masses.

The theory proposed in this letter can be considered as the 
renormalizable version of the theory given in reference~\cite{Borut-Goran}.
As we have discussed in this letter, the predictions coming from proton decay, 
the unification constraints and leptogenesis are quite different 
in this case. Our theory can be considered as the simplest 
renormalizable grand unified theory based on the SU(5) gauge symmetry.  
      
{\section{IV. Summary}}

We have investigated the possibility to find the simplest renormalizable 
grand unified theory based on the SU(5) gauge symmetry. We find that 
it is possible to generate all fermion masses with only 
two Higgs bosons, $5_H$ and $45_H$. In this context the neutrino masses 
are generated through the type III and type I seesaw mechanisms.
The predictions coming from the unification of gauge couplings and 
the stability of the proton have been discussed in detail. In this 
theory the leptogenesis mechanism can be realized through the out of 
equilibrium decays of the fermions $\rho_3$ and $\rho_0$ in the adjoint 
representation. We refer to this theory as ``Renormalizable Adjoint SU(5)".
\\
\\
\textit{Acknowledgments.}
{\small I would like to thank I.~Dorsner, R. Gonz\'alez Felipe and 
P.~Nath for the reading of the manuscript and useful comments. I would 
also like to thank B. Bajc and G. Senjanovi\'c for discussions and  
G.~Walsch for strong support. This work has been supported 
by { Funda\c{c}\~{a}o para a Ci\^{e}ncia e a Tecnologia} 
(FCT, Portugal) through the project CFTP, POCTI-SFA-2-777 and a
fellowship under project POCTI/FNU/44409/2002. I would like to thank 
the Max Planck Institut f\"ur Physik (Werner-Heisenberg-Institut) 
in M\"unchen for support and warm hospitality.}


\end{document}